# High-resolution disruption halo current measurements using Langmuir probes in Alcator C-Mod

**RA Tinguely[1],*, RS Granetz[1], A Berg[2], AQ Kuang[1], D Brunner[1], and B LaBombard[1]**
[1]MIT Plasma Science and Fusion Center, Cambridge, MA, USA
[2]Research Science Institute, Center for Excellence in Education, McLean, VA, USA

**Abstract.** Halo currents generated during disruptions on Alcator C-Mod have been measured with Langmuir "rail" probes. These rail probes are embedded in a lower outboard divertor module in a closely-spaced vertical (poloidal) array. The dense array provides detailed resolution of the spatial dependence (~1 cm spacing) of the halo current distribution in the plasma scrape-off region with high time resolution (400 kHz digitization rate). As the plasma limits on the outboard divertor plate, the contact point is clearly discernible in the halo current data (as an inversion of current) and moves vertically down the divertor plate on many disruptions. These data are consistent with filament reconstructions of the plasma boundary, from which the edge safety factor of the disrupting plasma can be calculated. Additionally, the halo current "footprint" on the divertor plate is obtained and related to the halo flux width. The voltage driving halo current and the effective resistance of the plasma region through which the halo current flows to reach the probes are also investigated. Estimations of the sheath resistance and halo region resistivity and temperature are given. This information could prove useful for modeling halo current dynamics.

## 1. Introduction

Diverted plasmas are vertically unstable and can move up or down during disruptions until eventually coming into contact with the first wall at one or more locations. Upon contact, the plasma converts to a limited configuration and continues to decay. In the relatively cold, resistive scrape-off layer (SOL) region formed by the limiter contact, "halo" currents can flow. The halo current circuit path through the SOL encircles the disrupting plasma core and is closed by passing through the first wall structure around the contact point or between multiple contact points. The changing poloidal and toroidal magnetic fluxes – from decaying plasma current and decreasing poloidal cross-sectional area, respectively – induce halo driving voltages substantially greater than the steady-state loop voltage needed to sustain a stable tokamak plasma. As a result, large $J \times B$ forces can arise in the structure, which is why halo currents are of concern during disruptions.

Measurements of halo currents have been made on a number of tokamaks for many years, traditionally using Rogowski coil sensors and/or resistive shunts [1-6]. The toroidal variation and rotation of the halo current distribution has been well-documented [3-7]. The poloidal structure has also been studied [3,5,6,8], but to a lesser degree and not as well-resolved, partly due to constraints on the spatial density of sensors. More detailed poloidal measurements would aid modeling work on halo currents, giving critical information about the halo channel width, resistivity (i.e. electron temperature $T_e$ and $Z_{eff}$) in the SOL, and the edge safety factor. In this paper we use a new dense array of Langmuir probes with high poloidal resolution on Alcator C-Mod [9,10] configured to measure halo current spatial distributions, voltages, and effective resistances.

In Section 2, we describe the experimental configuration and operation of the rail Langmuir probes for halo current measurements. The poloidal profile of halo current and magnetic reconstruction of the plasma boundary for a typical Vertical Displacement Event (VDE) are detailed in Section 3. In Section 4, the edge safety factor of the disrupting plasma is calculated for a variety of plasma discharges, and the halo flux width is approximated in Section 5. In Section 6, we measure the effective resistance of the halo region from data of six reproducible VDEs and estimate the resistivity and temperature of the halo region. Finally, Section 7 presents a summary of this work.

---

*Author to whom correspondence should be addressed: <rating@mit.edu>.

## 2. Rail Langmuir probes

Alcator C-Mod has extensive arrays of Langmuir probes embedded in its divertor structures to measure $T_e(\psi,t)$, electron density $n_e(\psi,t)$, and plasma potential $V_p(\psi,t)$ using standard analyses of I-V characteristics [11]. Typical Langmuir probes are small (~1-2 mm diameter) with domed, ramped, or flush surface geometries. In 2015, a new array of toroidally-extended ("rail" shaped), flush-mounted, field-aligned Langmuir probes was installed in a lower outboard divertor module (see Figure 1) [9,10]. This geometry was motivated by concerns of surviving reactor-level heat fluxes (with surface heat fluxes up to 100 MW/m$^2$) while minimizing sheath expansion effects. The array consists of 21 rail probes (19 of which were operational during the study described in this paper) arranged vertically along the face of one divertor module. The key features of this probe array important to this study are the small width of the probes (1.5 mm in the poloidal direction), long parallel length (~64 mm, which enhances current collection), and close spacing between probes (~1 cm, as shown in Figure 2). This is to be compared to the poloidal array of only four segmented Rogowski coils used for previous halo current measurements on C-Mod [3]. These were positioned 10 cm apart from the midplane downward, with only the bottommost measuring significant signal (see Figures 5(c) and 13 in Ref. [3]).

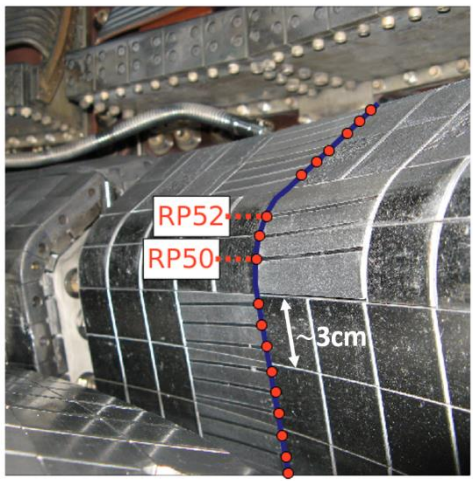

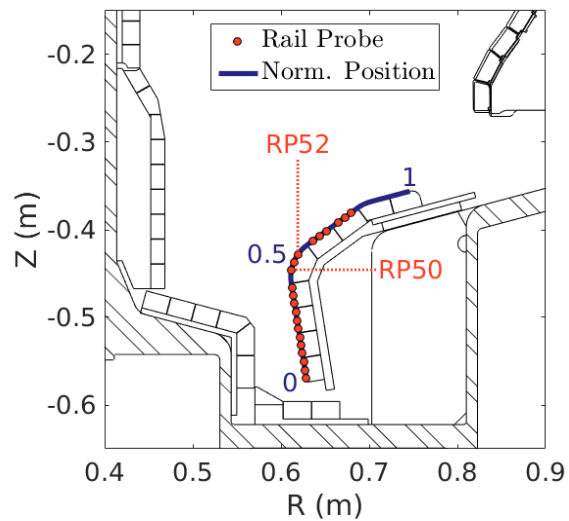

**Figure 1.** The rail probes are toroidally-elongated and field-aligned at 21 vertical positions along the outboard divertor. Red circles denote the poloidal locations of rail probes, with rail probes RP50 and RP52 labeled. Note that RP50-RP52 are shifted in toroidal angle due to space constraints of internal components.

**Figure 2.** A poloidal cross-section of the vacuum vessel shows the radial and vertical positions of 21 rail probes, marked by red circles (compare to Figure 1). The blue line indicates the normalized position along the divertor module face, which is used in this analysis. (The absolute arc length is ~30 cm.)

As with standard Langmuir probes, the rail probes are usually connected to power supplies that sweep their voltage with respect to the vacuum vessel (ground) repetitively and measure the collected current to obtain I-V characteristic curves, from which $T_e(\psi,t)$, $n_e(\psi,t)$, and $V_p(\psi,t)$ are extracted. The rail probes can also be run in "grounded" mode to mimic the surface of the divertor plate in which they are embedded. In this mode, the rail probes are not biased, but rather connected by a small resistor to ground. The probes collect current driven by electric fields produced by plasma dynamics, much as the surrounding divertor plate does, including halo current driven during disruptions. Since I-V characteristic curves are not obtained while running in grounded mode, it is typically employed only for fluctuations and disruption studies. The small width of the rail probes allows them to be packed into a relatively tight array in the poloidal, or $\psi$-coordinate direction; they provide a higher spatial resolution compared to the Rogowski sensors that were previously employed for measuring halo current distributions in C-Mod [3]. The rail probe current and voltage signals are digitized at 400 kHz, which provides good temporal resolution of the halo current dynamics during disruptions.

### 3. Halo current measurements

Reproducible downward VDEs were achieved on Alcator C-Mod by programming a downward "kick" in vertical position and subsequently turning off the feedback control. The plasma current and vertical position of an example VDE are shown in Figure 3, where the Current Quench (CQ) begins once the plasma has moved down far enough to become limited on the divertor plate. From the time the plasma becomes limited through the CQ (shaded in Figure 3), the rail probes measure halo current flowing into and out of the divertor tiles. Figure 4 displays the measured rail probe current density as a function of normalized divertor position (refer to Figure 2) and time. In these experiments, the plasma current and toroidal magnetic field were in the clockwise direction as viewed when looking down on the tokamak from above. Thus, halo current flows as indicated in Figure 5, i.e. into divertor tiles below the contact point and out of tiles above the contact point. The boundary between current entering the plate (positive, blue) and current exiting the plate (negative, red) tracks the motion of the plasma as it "slides" down the divertor plate, which is consistent with the vertical movement of the current centroid.

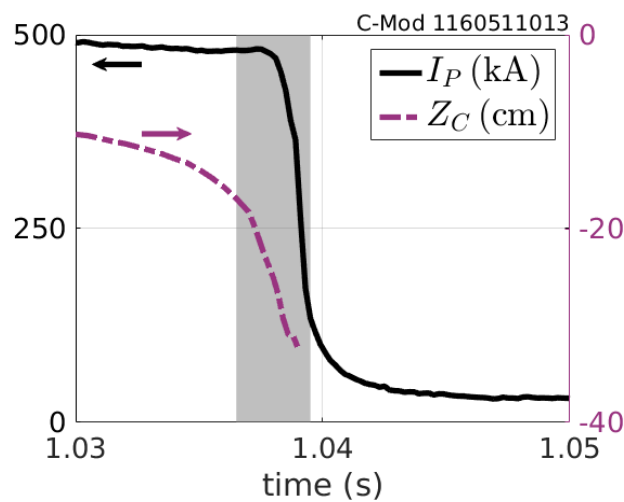

**Figure 3.** Plasma current ($I_P$, solid) and vertical position of the current centroid ($Z_C$, dash-dotted) are shown for a typical downward VDE. For this discharge, halo currents are observed in the shaded region (see Figure 4).

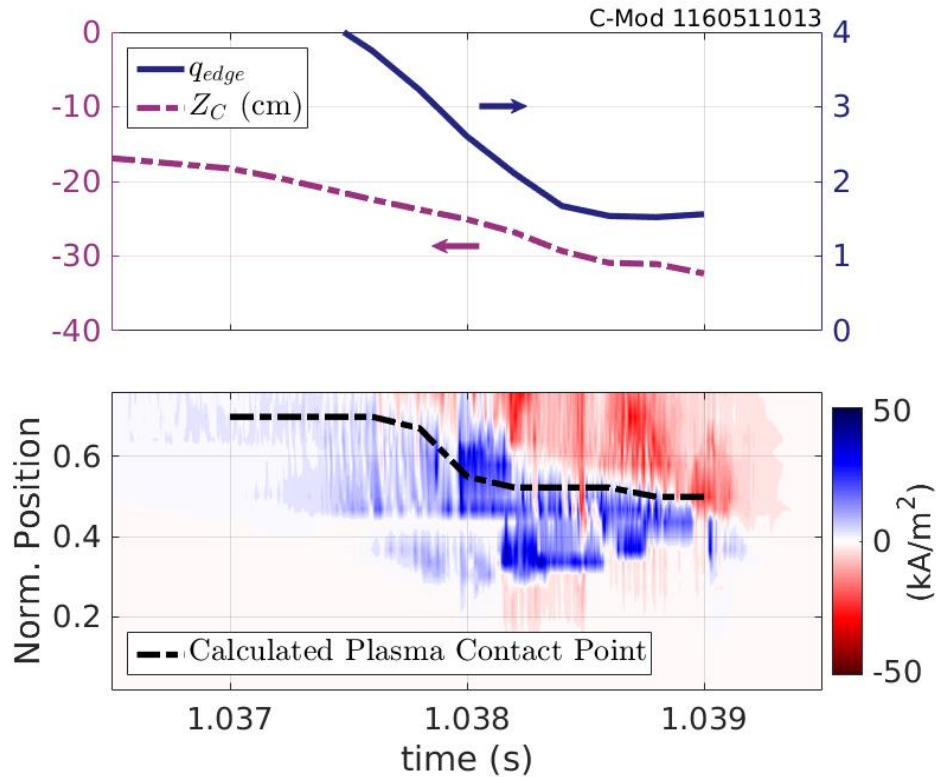

**Figure 4.** Top: Vertical position of the current centroid ($Z_C$, dash-dotted) and edge safety factor ($q_{edge}$, solid) are plotted for the same VDE as in Figure 3.

Bottom: Current density measurements from the rail probes vary over the normalized divertor position (see Figure 2) and time. Blue and red values of current density correspond to current flowing into and out of the rail probe, respectively. The calculated contact point (dashed line) of the plasma and divertor module is overlaid. Note that 20 Ω resistors were added to all rail probe circuits for this case; this reduces the magnitude of the halo currents collected by the probes, but is not expected to affect the poloidal profile shape – see Section 6. ($B_T$ = 5.4 T, $I_P$ = 0.5 MA)

During disruptions, standard Grad-Shafranov equation solvers (e.g. EFIT [12]) are not able to reconstruct the full magnetic geometry using external flux and field measurements. However, the vacuum magnetic field surrounding the disrupting plasma can be approximated by modeling the plasma current as an array of toroidal current-carrying filaments dispersed throughout the vacuum vessel, with the current distributed in a way that reproduces external magnetic measurements. A least-squares regression is found to be effective in determining the boundary of the plasma even during disruptions. Figure 5 displays the magnetic flux contours calculated by the filament reconstruction for several times during the downward VDE. The point of contact of the plasma with the divertor is calculated and is shown overlaying the rail probe current density measurements in Figure 4. Good agreement is seen between the calculated position and the measured location of the inversion from positive to negative current collection. For the particular discharge shown, the discrepancy between measured and calculated positions around t ~ 1.038 s could be explained by potential toroidal asymmetry and rotation of the falling plasma or the complicated motion of the plasma sliding over the divertor nose that is not adequately captured by the flux reconstruction.

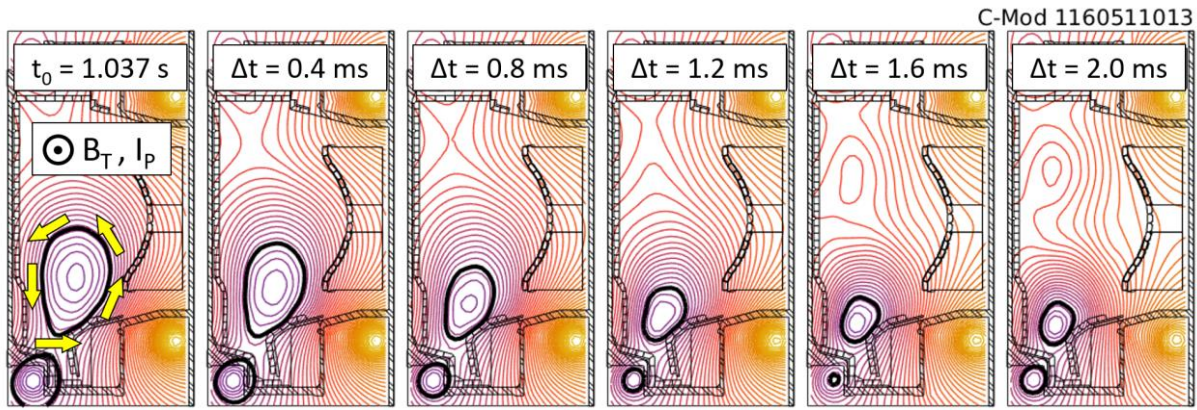

**Figure 5.** Poloidal magnetic flux contours are shown for several times, starting at $t_0 = 1.037$ s and in $\Delta t = 0.4$ ms increments, during a downward VDE. The bold black contours highlight the plasma boundary contacting and moving vertically down the lower outboard divertor. Arrows (yellow, leftmost panel) indicate the projection of halo current flowing through the SOL and the divertor module. The toroidal magnetic field ($B_T$) and plasma current ($I_P$) point in the direction shown.

## 4. Edge safety factor

The ability to reconstruct the vacuum magnetic topology during disruptions also provides the opportunity to calculate the safety factor at the edge of the disrupting plasma, $q_{edge}$. This information is important in simulations since certain events (namely the loss of thermal energy in the Thermal Quench, TQ) are triggered at given $q_{edge}$ values. However, in C-Mod, it is often impossible to identify the start of the TQ during VDEs, as the disruptions occur on timescales too short for EFIT reconstruction and out of sight of appropriate diagnostics (such as core $T_e$ measurements which do not view the divertor). In addition, few VDE discharges show the characteristic bump in plasma current associated with the TQ. Thus, it is of interest to explore the evolution of $q_{edge}$ from the start of the CQ through plasma termination.

The calculation of the edge safety factor proceeds as follows: During a VDE, the limiting plasma decreases both in minor radius and current. As $q_{edge} \propto a^2/I_P$, this typically results in a decreasing $q_{edge}$. At each point in time, the magnetic flux surface defining the boundary of the plasma is calculated from the filament reconstruction. At the divertor contact point, the system of 3-dimensional ordinary differential equations defining the magnetic field geometry is solved numerically using the Runge-Kutta fourth-order method to take incremental steps along the field line around the tokamak. Simply relating the number of toroidal revolutions per poloidal revolution gives the edge safety factor. The result of this calculation for our example VDE can be seen in Figure 4, where $q_{edge}$ decreases to ∼3/2 before final plasma termination.

In JT-60U, the TQ was found to occur when $q_{edge}$ was in the range 1.5-2 [13], and this value was subsequently used in simulation [14]. As mentioned, the TQ start time cannot usually be identified for VDEs in C-Mod; however, we have observed that the edge safety factor decreases in time before termination (see Figure 4). As a proxy, the minimum value of $q_{edge}$ is plotted in Figure 6 for plasma discharges at two magnetic fields ($B_T = 3.2, 5.4$ T) and three plasma currents ($I_P = 0.5, 0.7, 1.0$ MA). While the relationship between minimum edge safety factor and pre-disruption plasma parameters is not clear, many VDEs terminate at approximately rational minimum $q_{edge}$ values of ~1 and ~3/2. This motivates a wider range of $q_{edge}$ "triggers" to be explored in future simulations.

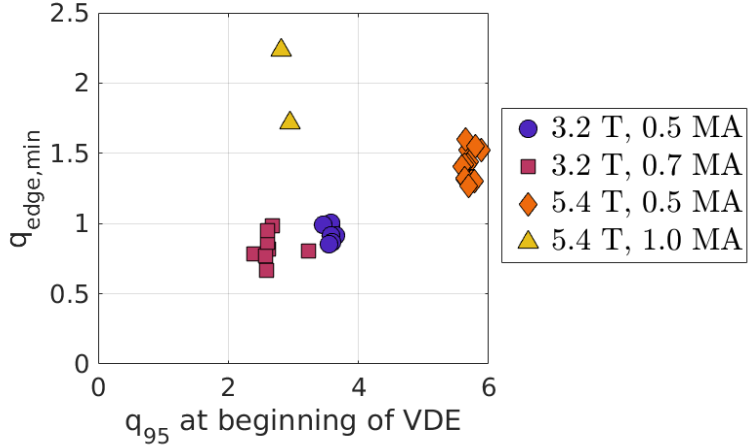

**Figure 6.** The minimum value of the edge safety factor $q_{edge,min}$ (calculated from flux reconstructions) is compared to the pre-disruption value of $q_{95}$ at the beginning of the VDE for $B_T$ = 3.2, 5.4 T and $I_P$ = 0.5, 0.7, 1.0 MA.

## 5. Halo flux width

The high spatial density of the rail probes is particularly well-suited for the measurement of the poloidal width of the halo current region. In order to increase current collection, the current-limiting resistors in the circuits for twelve rail probes around the divertor nose (with normalized positions of ~0.25-0.75) were set to 0 Ω. The halo current density measured by these probes for one discharge is shown in Figure 7. (Most VDEs show similar spatial and temporal patterns.) Again, the calculated contact point (dashed) from flux reconstructions matches the boundary between current flowing into and out of the rail probes and divertor plate. To determine the halo flux width, the lower bound of the positive current measurement (in blue) was mapped from the normalized divertor position (~0.35-0.4) to the flux surface intersecting the divertor at that point. The normalized flux width is calculated as

$$\Delta \overline{\psi} = \frac{\psi_{halo} - \psi_{edge}}{\psi_{edge} - \psi_{axis}} \qquad (1)$$

where $\psi_{halo}$, $\psi_{edge}$, and $\psi_{axis}$ are the values of poloidal flux at the boundary of the halo region, edge of the plasma (last closed flux surface as calculated by the filament reconstruction), and magnetic axis, respectively. It is important to note that sometimes the calculated halo flux contour will intersect the lower *inboard* divertor. Thus, the halo width given by Eq. (1) is actually a lower bound at these times. The total halo current can be approximated from the measured poloidal profile by assuming axisymmetry and integrating toroidally over the entire divertor surface. The time evolution of total halo current estimated from the rail probes is compared to that previously measured by Rogowski coils (from [3]) in Figure 8. Because total halo current values for these discharges are of the same order, we infer that the current that would be carried in this additional halo region is minimal, and therefore the lower bound of the halo width is a reasonable approximation.

For this downward VDE, the values of $\Delta \overline{\psi}$ varied in time from 0.15-0.6 (see Figure 9). The halo region is plotted for one time slice in Figure 7, where $\Delta \overline{\psi} \approx 0.43$. This is consistent with values used in simulations, such as in [15] in which $\Delta\overline{\psi}$ is defined as $\alpha_H$. Additionally, in Figure 7, the physical width of the outer contact (A-B) is larger than the inner contact (B-C). That is, due to the physical divertor geometry, the halo boundary contacts the outboard divertor plate farther above the plasma contact point than below it, which is consistent with the measured rail probe data.

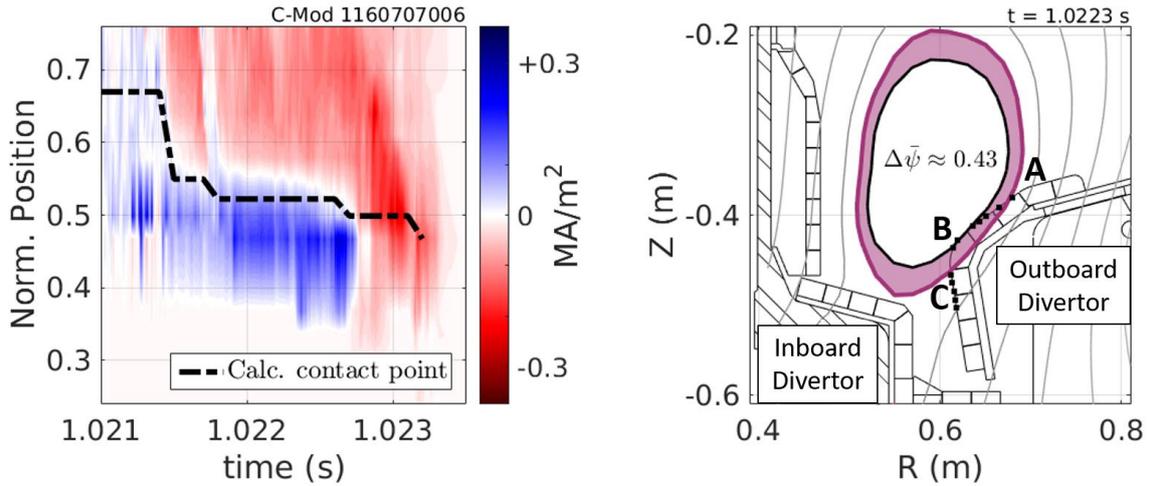

**Figure 7.** Left: Halo current density measurements from twelve grounded, shorted rail probes are plotted versus normalized divertor position and time. The calculated contact point from flux reconstruction is overlaid. ($B_T$ = 3.2 T, $I_P$ = 0.5 MA)

Right: Poloidal flux contours are shown for t = 1.0223 s with the plasma boundary in bolded black and the halo region shaded in purple. Points A and C indicate the intersections of the halo boundary and the divertor plate; point B is the plasma contact point. The twelve rail probes used are marked (black squares) along the divertor surface.

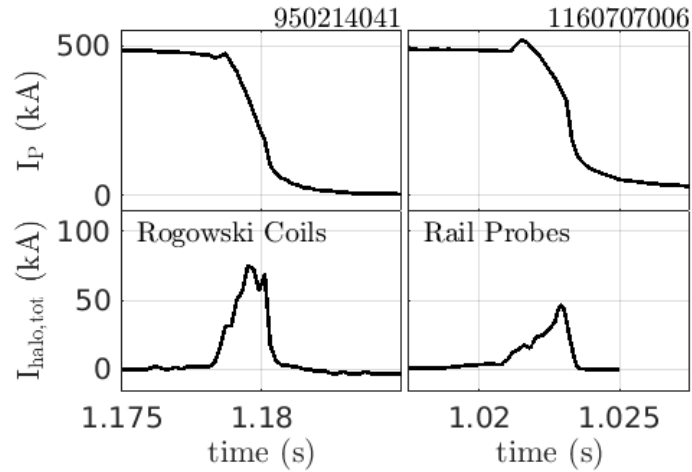

**Figure 8.** The plasma currents ($I_P$, top) and total halo current measurements ($I_{halo,tot}$, bottom) are shown for two similar VDEs, with $I_{halo,tot}$ measured by Rogowski coils and Rail Probes on the left and right, respectively. Both halo current measurements have similar time evolutions and magnitudes.

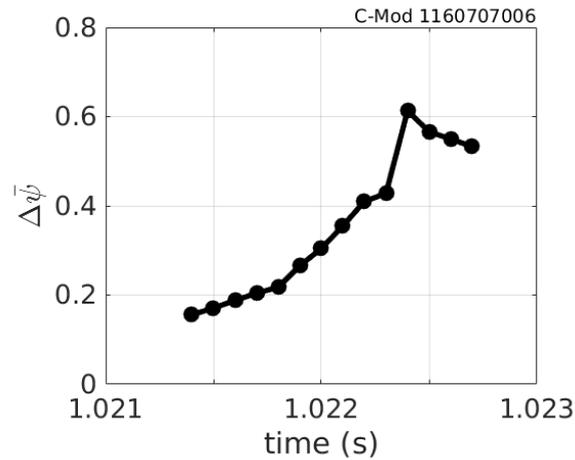

**Figure 9.** The normalized halo flux width increases in time for the VDE shown in Figure 7. Note that in general this calculation is a lower bound for the halo flux width.

## 6. Halo region resistance

As the falling plasma contacts the divertor surface, a closed circuit is formed from the open field lines in the boundary plasma through the divertor and vacuum vessel, with current flowing primarily along magnetic field lines in the plasma. The changes in poloidal and toroidal fluxes induce a voltage that drives current from the divertor, across a plasma sheath, through the halo flux region, across another sheath, and back into the divertor. The halo flux region can be idealized as a network of resistors parallel and perpendicular to the magnetic field lines. Assuming axisymmetry, there should be no net perpendicular current, and the resistor network acts as many halo flux tubes in parallel. Thus, if a rail probe were perfectly grounded (i.e. electrically shorted to the divertor module), we might imagine a halo current path as depicted in Figure 10.

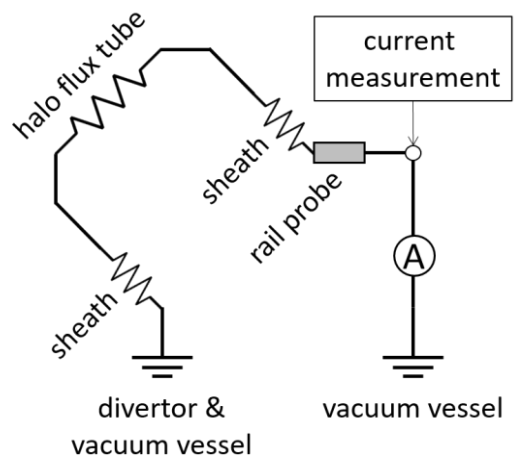

**Figure 10.** A schematic circuit diagram showing the ideal path of halo current through a perfectly-grounded rail probe. As the plasma limits on the divertor, an induced voltage (not depicted) in the halo region drives current through the flux tube incident on the rail probe, sheaths, and rail probe to ground.

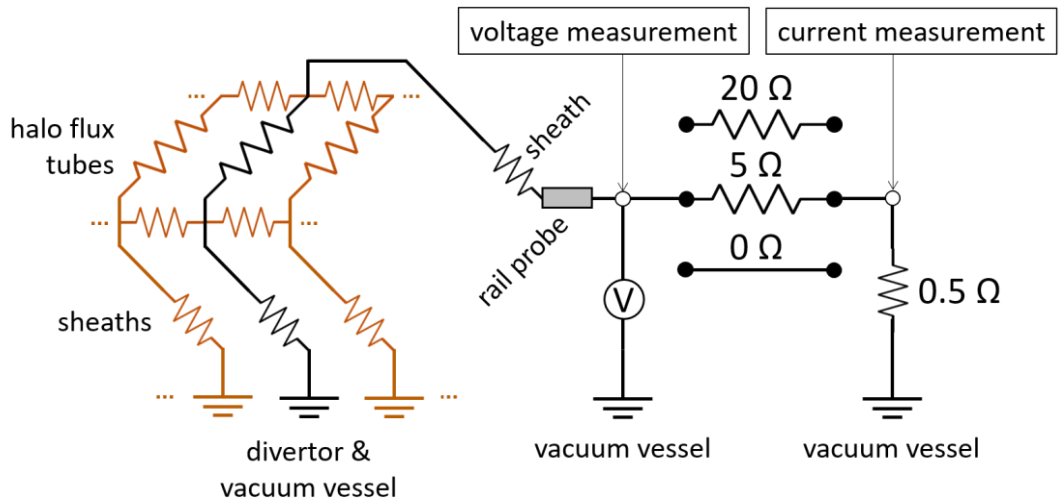

**Figure 11.** A schematic circuit diagram showing a more realistic halo current pathway through a rail probe. Due to the finite resistivity of the rail probe circuitry, the halo region acts as a resistor network with current spreading from the flux tube incident on the rail probe (shown in black) to adjacent flux tubes (shown in orange) and connecting to many points on the divertor and vacuum vessel. (See Figure 3 in Ref. [16] and compare to Figure 10 above.) Note that resistors can be added to this path (0, 5, and 20 Ω were available in this case) to change the amplitude of voltage and current measured while not affecting the plasma itself.

However, for halo current flowing into a rail probe, the physical picture is more complicated: Because the rail probe is connected to the vacuum vessel through a resistor, a voltage can develop between adjacent flux tubes causing halo current to spread between them due to the finite cross-field resistivity. Thus, the halo current loop closes by contact with many points on the vacuum vessel, as shown in Figure 11 and explored further in detail in Ref. [16]. While the relative levels of parallel versus cross-field resistivity are not discernible at this time, we can explore the effective resistance, $R_{eff}$, of the halo current pathway by changing the resistor in the probe circuit and noting the change in voltage and current measurements. This relies on two assumptions: first, the magnitude of the finite rail probe resistance does not affect the halo region resistor network, which is determined by plasma parameters like density and temperature; and second, halo current dynamics are similar for relatively reproducible VDEs, which were attained in experiment and are shown in Figure 12. As is discussed below, due to cross-field current sharing in the halo region, we find that this analysis primarily interrogates the sheath resistance and sets an upper limit on its value. In the end, this will only be a small correction to our estimation of the resistance of an ideal halo flux tube.

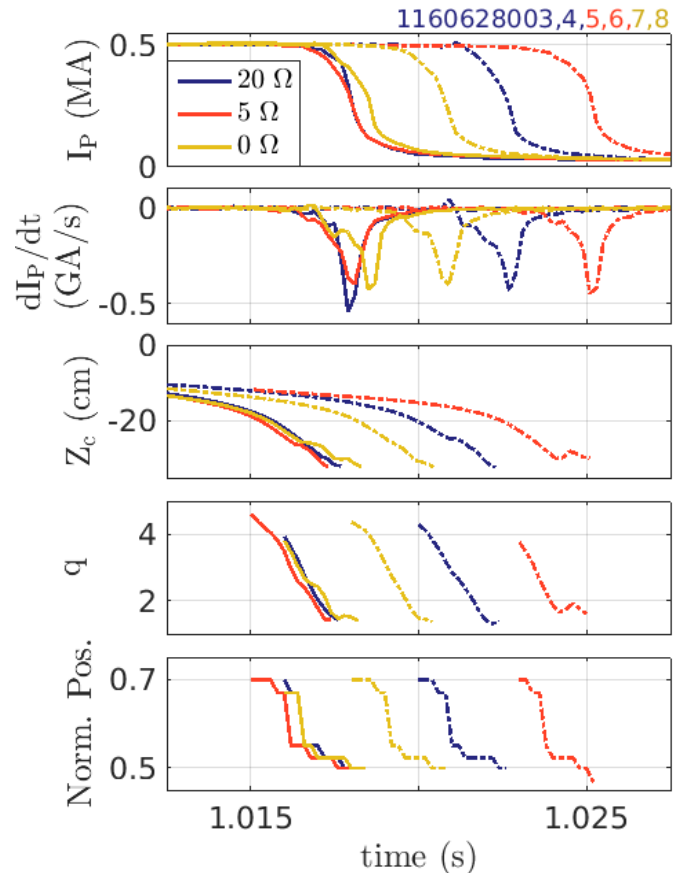

**Figure 12.** Plasma parameters are shown for six reproduced VDEs. From top to bottom: plasma current ($I_p$) and its rate of change ($dI_p/dt$), vertical position of the current centroid ($Z_c$), edge safety factor (q), and normalized position of the plasma contact with the divertor from filament reconstruction. Yellow, red, and blue colors correspond to the added resistor in the rail probe circuitry (0, 5, and 20 Ω, respectively), with dot-dashed lines representing the repeated discharge of the solid line with same color. Note that RP50 and RP52 have normalized positions of 0.44 and 0.5, respectively.

The resistance of each rail probe and the cable connecting it to the measurement electronics is ~0.2 Ω. An additional 0.5 Ω resistor is used for the current measurement. Another resistor is typically added in series to reduce the current and protect the diagnostic. In these experiments, this added resistance was varied among completely shorted (0 Ω), 5 Ω, and 20 Ω. The resulting currents and voltages measured by one rail probe with different added resistances are shown in Figure 13. As is seen, when the added resistor is increased from 0 to 5 to 20 Ω, the current measured through the rail probe decreases and voltage measured at the rail probe increases; thus, it is inferred that the effective resistance of the plasma region through which halo current flows is of the same order as the rail probe circuit resistances, i.e. around ~0.7-20.7 Ω. Also note that for the 0 Ω case, in which the rail probe measures the most current, the calculated total halo current is ~50-100 kA; this is on the same order as halo currents measured previously in C-Mod using passive Rogowski sensors which do not perturb the measured current (see Figure 8 and Ref. [3]).

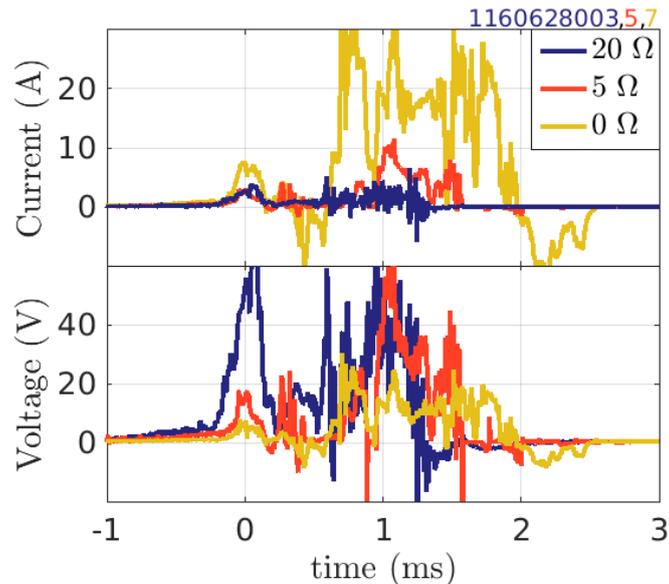

**Figure 13.** Current (top) and voltage (bottom) measurements for one rail probe (RP50) are shown for three different added resistances in the probe circuitry: 20 Ω (blue), 5 Ω (red), and 0 Ω (yellow). Note that the time base of each discharge has been shifted so that the initial peak in signal occurs at t ~ 0 ms.

Due to the limited number of extra rail probe circuit cards, only four probes were used in this experiment, and of those only two (RP50 and RP52) had appropriate signal levels to generate useful data for this particular analysis. From the bottom plot of Figure 12, note that the normalized positions of RP50 and RP52 – 0.44 and 0.5, respectively – are almost always below the plasma contact point and therefore primarily measure positive current (ions flowing into the rail probe). It is important to note that the effective resistance deduced from positive measured current may differ from one deduced from a negative measured current. In fact, we expect the effective resistance deduced from a negative measured current to be lower; in regimes in which a probe collects a negative current, the electron collection increases as ~exp(eV/kT$_e$) [11], i.e. the effective plasma sheath impedance is reduced. Thus, for the analysis below, we consider our effective resistance to be an upper bound of the actual value.

The data in Figure 13 show three characteristic features: an initial positive peak in signal followed by a broader positive feature and ending with a negative bump. However, due to the different timings of these features, it is difficult to directly compare the time evolution of data from different discharges. Instead, the data were averaged over in time over the broad positive feature to identify one voltage and current measurement for each discharge. The basic circuit equation is

$$V_{eff} = I_{measured} \times R_{eff} + V_{measured} - L_h \frac{d}{dt}(I_{measured}) \qquad (2)$$

where $V_{eff}$ is the voltage required to drive current through the effective resistance of the halo path $R_{eff}$, $I_{measured}$ and $V_{measured}$ are rail probe measurements, and $L_h$ is the inductance of the halo current path. Due to the rapid variation of $I_{measured}$ in time, the voltage arising from the inductance averages to zero; even assuming a value of $L_h$ similar to the plasma (~1 μH) results in peak voltages on the order of 1 V or less, so the last term can be neglected. For a single discharge, we only know that $V_{eff}$ and $R_{eff}$ follow this linear relationship, but not their precise values; in other words, $V_{eff}$ and $R_{eff}$ can lie anywhere on a line specified by $I_{measured}$ and $V_{measured}$. Exactly-reproduced discharges (with the same rail probe circuitry) should generate the same (i.e. overlaying) lines. For multiple reproducible discharges with *different* rail probe circuit resistances – and thus different values for $I_{measured}$ and $V_{measured}$, the multiple lines should intersect at the actual $V_{eff}$ and $R_{eff}$.

Following this reasoning, Figure 14 shows the linear relationships from Eq. (2) for six reproduced VDEs ($B_T = 5.4$ T, $I_P = 0.5$ MA, $\bar{n}_e \sim 1.2 \times 10^{20}$ m$^{-3}$). There are two discharges (solid and dashed lines) for each of the three possible added resistances to the probe circuit (denoted by different colors). The consistency in slope and y-intercept (average measured current and voltage, respectively) for the same added resistance gives us confidence in the reproducibility of these VDEs. Note that the intersections of all six lines occur around $R_{eff} \sim 0.5\text{-}2$ $\Omega$, meaning that the actual effective resistance of the halo region lies approximately in that range.

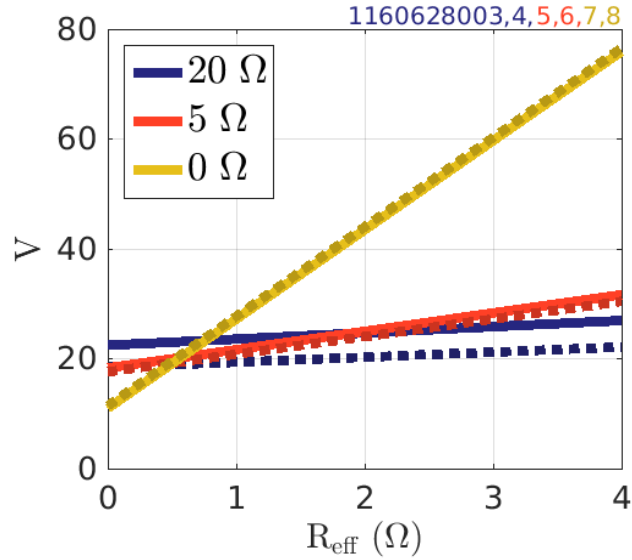

**Figure 14.** Driving voltage (for RP50) is plotted over a range of possible effective resistances satisfying Eq. (2) for six reproduced VDEs. Two discharges (solid and dashed lines) were run for each added resistance to the rail probe circuitry (0, 5, and 20 $\Omega$).

Because the VDEs were reproducible (refer to Figure 12), we assume that the voltage driving halo current to the rail probe is the same for each VDE (refer to Table 2). Thus, the effective halo resistance, $R_{eff}$, is calculated from the system of equations representing the intersections of these lines:

$$R_{eff} = -(V_{m,i} - V_{m,j})/(I_{m,i} - I_{m,j}) \qquad (3)$$

where $V_m$ and $I_m$ are the measured voltages and currents from two discharges (*i* and *j*) with different added resistances. The resulting mean and standard deviation of the calculated resistance for two rail probes is given in Table 1. While time-averaged values of measured current and voltage were used in their calculation, these mean resistances can be plugged back into Eq. (2) to show similar evolutions of $V_{eff}$ in time, which is seen in Figure 15.

**Table 1.** Mean resistance ($\pm$ standard deviation) in the halo region for rail probes RP50 and RP52 (see Figure 1 and Figure 2).

| Rail Probe | Resistance $R_{eff}$ ($\Omega$) |
|---|---|
| RP50 | $0.8 \pm 0.7$ |
| RP52 | $1.3 \pm 0.9$ |

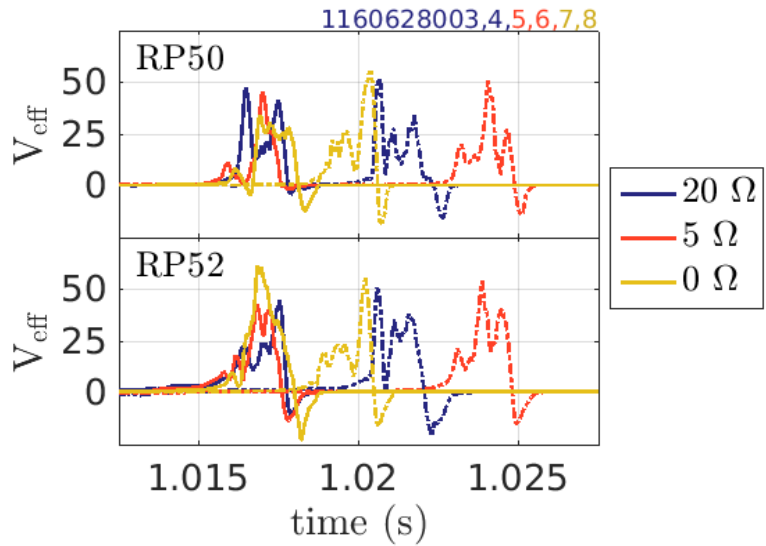

**Figure 15.** The time evolutions of Eq. (2) are plotted for rail probes RP50 and RP52 using mean $R_{eff}$ values from Table 1, measured currents and voltages, and $L_h = 1$ µH. The six traces correspond to the same discharges as in Figure 12, and a 0.25 ms smoothing window was applied to the data to highlight similar features.

The electric field driving halo current has both poloidal and toroidal components resulting from the respective change in magnetic fluxes. Specifically, the poloidal electric field is induced by the collapsing plasma cross-section in the background toroidal magnetic field (approximately constant in time), and the toroidal electric field is generated by the decay of the total plasma current. By finding the projection of the electric field along the magnetic field line (to which current flows nearly parallel in the halo region), an induced voltage $V_{loop}$ can be calculated as in [4]:

$$V_{loop} = V_{pol} + q \cdot V_{tor} \approx -\left(B_0 \frac{dA}{dt} + q \cdot L \frac{dI_P}{dt}\right) \quad (4)$$

where $V_{pol}$ and $V_{tor}$ are poloidal and toroidal voltages, respectively; q is the safety factor; $B_0$ is the toroidal magnetic field on axis; A is the plasma poloidal cross-sectional area; L is the plasma inductance (~1 µH in C-Mod); and $I_P$ is the total plasma current.

For the six discharges studied, $V_{pol}$ and $V_{tor}$ can be evaluated as functions of time, where $B_0$ and $I_P$ are measured experimentally, and q and A are calculated from flux reconstructions. As seen in Table 2, these voltages are found to be of similar order, and averaged-in-time the total $V_{loop}$ ranges from ~210-250 V. From this, the resistance of the halo current path can be estimated from the shorted rail probe case – which most closely resembles an electrical short – to be ~13.7 Ω (where $V_{loop}$ ~ 230 V from the average of the bottom two rows of Table 2 and $I_{measured}$ ~ 16 A and $V_{measured}$ ~ 11 V from the yellow signals for RP50 in Figure 13). The discrepancy between this resistance and that measured above seems to indicate that cross-field current flow dominates at the rail probe and that the measured $R_{eff}$ ~ 1 Ω is an upper bound of the rail probe sheath resistance. In other words, the resistance of the rail probe compared to the surrounding divertor enhances current-sharing between adjacent flux tubes, as shown in Figure 11; thus, in the limit of complete cross-field current flow, our analysis interrogates only the sheath resistance at the rail probe.

**Table 2.** Poloidal, toroidal, and total loop voltages ($V_{pol}$, $V_{tor}$, and $V_{loop}$, respectively) from Eq. (4), are averaged in time for six reproduced plasma discharges and three different resistors added to the rail probe circuitry. Note that the values of $B_0$ = 6 T and L = 1 μH were taken to be constant in time.

| Discharge No. | Rail Probe Resistor (Ω) | $V_{pol}$ (V) | $V_{tor}$ (V) | $V_{loop}$ (V) |
|---|---|---|---|---|
| 1160628003 | 20 | 165 | 86 | 251 |
| 1160628004 | 20 | 113 | 98 | 211 |
| 1160628005 | 5 | 99 | 139 | 238 |
| 1160628006 | 5 | 83 | 168 | 251 |
| 1160628007 | 0 | 102 | 115 | 217 |
| 1160628008 | 0 | 112 | 133 | 245 |

This information also provides an opportunity to estimate the resistivity of the halo flux region. For the purpose of this calculation, we assume that a truly-shorted rail probe would measure the same halo current as the almost-shorted case detailed above. Then, the rail probe would behave just like the surrounding divertor surface, and the halo current pathway would most resemble the halo flux tube in Figure 10. The resistivity calculation is

$$\eta_{halo} = R_{halo} A_{cs} / L_c \quad (5)$$

where $R_{halo}$ ~ 12.1 Ω is the total halo resistance *minus* ~1.6 Ω (which is an upper limit of the contribution from two sheaths), $A_{cs}$ is the cross-sectional area of the flux tube incident on the rail probe, and $L_c$ is the connection length of the flux tube. The cross-sectional area depends on the rail probe surface normal (measured in [10]) and magnetic field vector calculated from the filament reconstruction; the time- and discharge-averaged value for these discharges is $A_{cs}$ ~ 5 mm$^2$, which corresponds to an angle of incidence between the surface normal and field line of ~87 degrees. While the edge safety factor changes during the disruption, this calculation uses an averaged value of $q_{edge}$ = 2.25, and $L_c$ ~ $2\pi R_0 q_{edge}$, where $R_0$ = 0.68 m is the major radius of C-Mod.

The estimated halo region resistivity is $\eta_{halo} \approx 6.3$ μΩ-m, which results in temperatures (from Spitzer [17,18]) of $T_e$ ~ 35-58 eV for $Z_{eff}$ = 1-2. Note that including the sheath resistance in the calculation only changes $T_e$ by <10%. The lower end of this $T_e$ estimate is consistent with common simulation values of halo region temperature, which typically range from 1-50 eV for $Z_{eff}$ = 1-2. The relatively high temperatures at the upper end of our estimate could be due to a TQ late in the CQ, so the plasma may still be relatively hot at the time of halo current measurements. In addition, these probes are positioned directly on the nose of the divertor (see Figure 1 and Figure 2), meaning they are often closest to the limiting point of the plasma (see Figure 12) and thus could experience higher temperatures. Furthermore, this information could be useful for implementation of boundary layer physics and even synthetic diagnostics in future VDE/halo current simulations.

Finally, a simple power balance analysis can be performed. For the high voltages induced during the VDE, we could presume that the halo current measured by a grounded rail probe is on the order of the ion saturation current. Using Equation (25.43) and a sheath transmission coefficient of ~7 from Figure 25.10 in Ref. [19] and the values of $I_{measured}$ and $A_{cs}$ from above, the resulting heat flux *parallel* to the magnetic field is ~0.8-1.3 GW/m$^2$ for $T_e$ ~ 35-58 eV. Taking into account the angle of incidence of the field-line on the divertor surface reduces this heat flux to 41-68 MW/m$^2$. The energy lost by the plasma is dominated by the magnetic energy, which is usually an order of magnitude larger than the stored thermal energy. For these discharges, the average power lost from the decay of the plasma current during the time of halo current measurements is ~30 MW. Assuming axisymmetry and that the heat flux estimation is peaked around the divertor nose, the power lost from the plasma to the divertor is consistent.

## 7. Summary

A vertical array of closely-positioned, flush-mounted, toroidally-extended, and field-aligned Langmuir "rail" probes along the lower outboard divertor has recently been used to measure the halo current profile resulting from downward VDEs in Alcator C-Mod. When grounded to the vacuum vessel, including no added resistance in their circuitry, the rail probes measure total halo currents on the same order as those measured previously in C-Mod with Rogowski coils, but with a finer poloidal resolution. The contact point of the limited plasma with the divertor is clearly seen in the data and is consistent with magnetic flux reconstructions of the disruption. From these reconstructions, the edge safety factor is computed, and the minimum value is observed to vary with magnetic field and pre-disruption plasma current, but remains at approximately rational values of 1 and 3/2. The halo region width is also measured to vary between ~15-60% of the normalized poloidal flux. Finally, an upper bound on the sheath resistance is determined from six reproduced VDEs, and the halo region resistivity and temperature are estimated to be ~6.3 μΩ-m and ~35-58 eV, respectively. These measurements motivate further experimental investigation using poloidal probe arrays on other devices. In addition, these results could guide future simulation efforts, which require such inputs as edge safety factor and halo region width and resistance/resistivity.

**Acknowledgments**
This work supported by US DOE Grant DE-FC02-99ER54512, using Alcator C-Mod, a DOE Office of Science User Facility.